\documentclass[fleqn,12pt]{JHEP3}
\usepackage{amsmath}
\usepackage{epsfig}
\usepackage{amsfonts,amssymb,latexsym}
\usepackage[nosort]{cite}

\ifx\href\asklfhas\newcommand{\href}[2]{#2}\fi

\newdimen\squaresize \squaresize=12pt
\newdimen\thickness \thickness=0.7pt

\def\square#1{\hbox{\vrule width \thickness
   \vbox to \squaresize{\hrule height \thickness\vss
      \hbox to \squaresize{\hss#1\hss}
   \vss\hrule height\thickness}
\unskip\vrule width \thickness} \kern-\thickness}

\def\cut#1{\hbox{\vrule width-1 \thickness
   \vbox to \squaresize{\hrule height-1 \thickness\vss
      \hbox to \squaresize{\hss#1\hss}
   \vss\hrule height-1\thickness}
\unskip\vrule width +4 \thickness} \kern-\thickness}

\def\vsquare#1{\vbox{\square{$#1$}}\kern-\thickness}

\makeatletter
\newbox\slashbox \setbox\slashbox=\hbox{$/$}
\def\pFMslash#1{\setbox\@tempboxa=\hbox{$#1$}
  \@tempdima=0.5\wd\slashbox \advance\@tempdima 0.5\wd\@tempboxa
  \copy\slashbox \kern-\@tempdima \box\@tempboxa}

\makeatother

\newcommand{\w}{\omega}

\newcommand{\ft}[2]{{\textstyle {\frac{#1}{#2}} }}

\newcommand{\be}{\begin{equation}}
\newcommand{\ee}{\end{equation}}
\newcommand{\ben}{\begin{displaymath}}
\newcommand{\een}{\end{displaymath}}
\newcommand{\ba}{\begin{eqnarray}}
\newcommand{\ea}{\end{eqnarray}}
\newcommand{\nn}{\nonumber}

\newcommand{\bean}{\begin{eqnarray*}}
\newcommand{\eean}{\end{eqnarray*}}

\def\moth{\mathsurround=0pt}
\newdimen\zo \zo=0pt

\def\tick{\leaders\hrule height 0.5ex depth 0pt \hskip 0.5pt}
\def\upboxfill{$\moth \setbox\zo\hbox{\tick}%
  \hskip 2pt\hbox to 0pt{$\tick$\hss}\hrulefill \hbox to
6pt{$\tick$\hss}$}

\def\dtick{\leaders\hrule height .34pt depth .5ex \hskip 0.5pt}
\def\downboxfill{$\moth \setbox\zo\hbox{\dtick}%
  \hskip 2pt\hbox to 0pt{$\dtick$\hss}\hrulefill \hbox to
6pt{$\dtick$\hss}$}

\newcommand{\Ref}[1]{(\ref{#1})}

\makeatletter \@addtoreset{equation}{section} \makeatother

\renewcommand{\thetable}{\Roman{table}}

\def\be{\begin{equation}}
\def\ee{\end{equation}}
\def\bea{\begin{eqnarray}}
\def\eea{\end{eqnarray}}

\title{  Entropy function and attractors for AdS black holes }

\author{  Jose F. Morales\\
CERN, Theory Division, CH-1211, Geneva 23, Switzerland\\
INFN, Universit\`a di Roma ``Tor Vergata'', 00133 Rome, Italy\\
E-mail: \email{Francisco.Morales.Morera@cern.ch} }

\author{  Henning Samtleben\\
Zentrum f\"ur Mathematische Physik, Universit\"at Hamburg\\
Bundesstrasse 55, D-20146 Hamburg, Germany \\
 E-mail: \email{Henning.Samtleben@desy.de} }

\abstract{ We apply Sen's entropy formalism to the study of the
near horizon geometry and the entropy of asymptotically AdS black
holes in gauged supergravities. In particular, we consider
non-supersymmetric electrically charged black holes with
$AdS_2\times S^{d-2}$ horizons in $U(1)^4$ and $U(1)^3$ gauged
supergravities in $d=4$ and $d=5$ dimensions, respectively. We
study several cases including static/rotating, BPS and non-BPS
black holes in Einstein as well as in Gauss-Bonnet gravity. In all
examples, the near horizon geometry and black hole entropy are
derived by extremizing the entropy function and are given entirely
in terms of the gauge coupling, the electric charges and the
angular momentum of the black hole. }


\preprint{CERN-PH-TH/2006-158}

\begin{document}

\tableofcontents

\bigskip
\bigskip

\section{Introduction}

The study of black hole thermodynamics has played a central role
in the development of our current notions of holography in gravity.
In this line of thinking, black holes are viewed as thermodynamic
objects at equilibrium with a temperature and an entropy. A simple analysis
of this thermodynamic system leads to the remarkable  Bekenstein-Hawking formula for
the black hole entropy. This formula relates the entropy of the black hole to
the area of its horizon and it suggests that the microscopic
degrees of freedom of the black hole can be described by a ``dual" quantum mechanics
living on the horizon. This is further supported by the discovery of
AdS/CFT dualities \cite{Maldacena:1997re} that relate gravity on AdS spaces and gauge theories living on the AdS boundary.     These observations drastically simplify the study of black hole
physics, since the geometry of the horizon is typically much simpler than that of
the full solution. Even in theories with scalar fields and a large number of moduli
-- asymptotic values of massless scalars at infinity --, scalars are attracted
at the black hole horizon to special values and the full geometry is entirely determined
in terms of the black hole charges.
This is referred as
the {\it attractor mechanism}~\cite{Ferrara:1995ih,Strominger:1996kf,Ferrara:1996dd,Ferrara:1996um}.
Originally discussed in the context of ${\cal N}=2$ black holes the attractor mechanism has been
recently extended in many directions, including non-supersymmetric and higher
derivative gravity theories~\cite{Goldstein:2005hq,Kallosh:2005ax,Tripathy:2005qp,Giryavets:2005nf,Goldstein:2005rr,Kallosh:2006bt,Chandrasekhar:2006kx,Bellucci:2006ew,Kallosh:2006bx,Ferrara:2006em,Ferrara:2006xx,Bellucci:2006xz,Ferrara:2006yb,Kallosh:2006ib}.
The results show that the attractor mechanism is a universal issue of any gravity theory.

In \cite{Sen:2005wa}, A. Sen introduced a unifying formalism, the {\it entropy formalism},
that  describes the attractor equations and black hole entropy in a general non-supersymmetric
and higher derivative gravity theory.
In this formalism, the near horizon geometry is determined by extremizing a single
function $F$, the {\it entropy function}. The entropy of the black hole is given
by the value of $F$ at the extremum. The function $F$ is defined by
the Legendre transform with respect to the black hole charges
of the gravity action evaluated at the horizon.
More precisely, the gravity action is first evaluated at a trial background
geometry with volumes and scalar/gauge field profiles parametrized
by a finite number of parameters. These parameters are then determined by extremizing the entropy
function $F$.
The formalism has been successfully applied to the study of general non-supersymmetric
asymptotically flat black holes in various supergravity settings
 \cite{Sen:2005iz,Alishahiha:2006ke,Sahoo:2006rp,Cardoso:2006nt,Chandrasekhar:2006zw,Ghodsi:2006cd,Alishahiha:2006jd,Cai:2006xm,Astefanesei:2006dd,Cardoso:2006cb,k1,k2}.

 The aim of this paper is to extend this analysis to the study of
 asymptotically AdS black holes in gauged supergravities.
 According to holography \cite{Maldacena:1997re} the entropy of black holes in AdS spaces
 is related to the free energy of the dual gauge theory living on the AdS boundary,
see \cite{Witten:1998zw,Sundborg:1999ue,Aharony:2003sx,Kinney:2005ej,Alvarez-Gaume:2005fv,Alvarez-Gaume:2006jg}.
 To pursue the study of these holographic correspondences a detailed knowledge
of the black hole near horizon data is required.
To derive explicit formulas for the attractor geometry and for the entropy of AdS black
holes is one of the main motivations of the present work.

Black holes in gauged supergravities are different from those in Poincar\'e supergravities
 in many respects. First, in the gauged theory the asymptotic values of the scalar
 fields at infinity are typically fixed at the minimum of a scalar potential.
 The moduli space is therefore reduced and often empty. Still once charges are placed
on $AdS_d$, even scalars fixed at infinity flow at the horizon to a different fixpoint specified
completely by the black hole charges. I.e. the attractor mechanism now describes a flow between
two fixpoint geometries. Second, it is well known that asymptotically AdS black hole solutions
with regular horizons are always non-supersymmetric unless a non-trivial
angular momentum is turned on. This is very different from the Minkowski case where
BPS static solutions are quite common. Our analysis here explores both
non-supersymmetric static and rotating black hole solutions.

We apply the entropy formalism to non-supersymmetric black holes with
 near horizon geometry $AdS_2\times S^{d-2}$ in $d=4,5$\footnote{
More precisely, in
 the case of rotating black holes the
horizons are described by a ``squashed $AdS_2\times S^{d-2}$'' rather than
a tensor product geometry.}.
Black holes with these type of horizons have always zero temparature (with coinciding inner and outer horizons)
but they are in general  non-supersymmetric.
For concreteness we focus on the $U(1)^4$ and $U(1)^3$ gauged supergravities in $d=4$ and $d=5$, respectively.
These theories can be embedded into the maximal gauged supergravities with
gauge groups $SO(8)$ and $SO(6)$, respectively, following from
compactifications of M-theory and type IIB theory on $AdS_4\times S^7$ and $AdS_5\times S^5$,
respectively. Black holes in these gauged supergravities have been extensively studied
and classified in full generality in the  literature
 \cite{Behrndt:1998jd,Cvetic:1999xp,Duff:1999gh,Sabra:1999ux,Gutowski:2004ez,Gutowski:2004yv,Cvetic:2004hs,Cvetic:2004ny,Cvetic:2005zi,Chong:2005da,Chong:2005hr,Kunduri:2006ek}
 (see \cite{Duff:1999rk} for a review and a list of references).
In the case of Einstein gravity, the solutions derived here via the entropy formalism
follow from these general solutions by taking the zero temperature limit.
Our focus here is on the near horizon geometry and black hole entropy.

We test the entropy formalism in a number of examples, including
static/rotating black holes  with or without supersymmetry in Einstein as
well as Gauss-Bonnet gravity.
In each case we show that the attractor geometry follows
from extremization of the entropy function. In the case of Einstein gravity the entropy
function output will be shown in agreement with the Bekenstein-Hawking formula as expected.

The entropy formalism is particularly efficient in the study of black holes in higher
derivative gravity.
Higher derivative corrections to black hole entropies
in rigid supergravities were first studied in
\cite{LopesCardoso:1998wt,LopesCardoso:1999ur,LopesCardoso:2000qm,Dabholkar:2004yr}.
Higher derivative corrections to asymptotically AdS black holes in
Gauss-Bonnet gravity were studied in \cite{Cvetic:2001bk}.
More recently in~\cite{Dutta:2006vs} the authors consider several examples of higher
derivative terms and derive the first corrections to the Schwarzschild  AdS
black holes. Here we consider
the Einstein-Maxwell system in the presence of a Gauss-Bonnet term and derive
exact expressions for the near horizon geometry and the black hole entropy.

The paper is organized as follows: In sections 2 and 3 we consider non-rotating
asymptotically AdS black holes
in $U(1)^4$ and $U(1)^3$ gauged supergravities in $d=4$ and $d=5$, respectively.
In section 4 we apply the entropy formalism
to rotating black holes in $d=5$ gauged supergravity.
The study of higher derivative corrections is
sketched in section~5 for the Gauss-Bonnet type of interactions in the Maxwell-Einstein
system in $d=4,5$ dimensions.
In Section 6 we summarize our results and draw some conclusions.
Appendix~\ref{units} contains a discussion on the normalization of the physical charges used in the main text.
Appendix \ref{scomparison} presents the link between our $AdS_2\times S^{d-2}$ solutions  and zero
temperature limits of the general black hole solutions.

\section{AdS$_4$ static black holes}
\label{nonrot4}

We start by considering $U(1)^4$ gauged supergravity in four dimensions.
This theory follows from a
truncation of the maximal $N=8$, $SO(8)$ gauged supergravity~\cite{deWit:1982ig} down to the
Cartan subgroup of $SO(8)$. The bosonic action can be written as~\cite{Duff:1999gh}:
\be
S={1\over 16 \pi G_4}\int d^4x \,\sqrt{-g}\,
\left[ R-\ft14 X^2_{I}
F_{\mu\nu}^I F^{\mu\nu I} -\ft12\, X_I^{-2}\,\partial_{\mu} X_I
\,\partial^{\mu} X_I - V \;\right] \;,
\ee
with $I=1, \dots, 4$, and
\be
F_{\mu\nu}^I=2\partial_{[\mu}A^{I}_{\nu]}\;,\qquad
V= -4\, g^2\, \sum_{I<J} X_I \, X_J \;,\qquad
X_1 X_2 X_3 X_4=1\;.
 \ee
The equations of motion derived from this lagrangian are:
\bea
&& R_{\mu \nu}-\ft12\, X_I^2\, F^I_{\mu \sigma}\, F_{\nu}^I {}^{\sigma}
-\ft12 X_I^{-2} \partial_\mu X_I \,\partial_\nu X_J
-\ft12\, g_{\mu\nu} \left(R-\ft14\, X^2_{I}\, F^{I2}-\ft12\, (X_I^{-1}\partial X_I)^2
-V \right)=0\;,\nn\\[1ex]
&& {\delta\over \delta X_I} \left(\ft14\, X_{I}^2\, F^{I2}+\ft12\, (X_I^{-1}\partial X_I)^2
+V\right)=0\;,\nn\\[1ex]
&&\partial_\mu(\sqrt{-g}\,X_{I}^2 \, F^{\mu \nu I})=0\;.\label{eom4}
\eea
We look for non-rotating black hole solutions with $AdS_2\times S^2$ near horizon
geometry
\bea
ds^2 &=& v_1\, \Big( -r^2\,dt^2+ {dr^2\over r^2}\Big) +
v_2\,d\Omega_2 \;,\nn\\
 X_{I} &=& u_I\;,\qquad A^I = -e_I \, r\, dt \;,\qquad
 F^I_{0r}= e_I\;,\nn\\[1ex]
d\Omega_2 &=&(d\theta^2+\sin^2\theta\, d\phi^2)\qquad 0\leq \theta\leq \pi\quad 0\leq \phi <2\pi
\;,
\label{ansatz0}
\eea
 with constants $u_I, e_I, v_a$,
 and $u_4=1/(u_1 u_2 u_3)$.

 The attractor equations determining the constants $u_I,v_a,e_I$
 at the black hole horizon are efficiently described by the so called
 entropy formalism \cite{Sen:2005wa}. One starts by evaluating the supergravity
 action (integrated on the $S^2$ horizon) in the background (\ref{ansatz0}):
\be
f(\vec{e},\vec{v},\vec{u}) \equiv
\int d\theta d\phi\, \sqrt{-g}\, {\cal L}(\vec{e},\vec{v},\vec{u})
\;,
\ee
with ${\cal L}(\vec{e},\vec{v},\vec{u})$ the Lagrangian density evaluated on the
ansatz (\ref{ansatz0}).
The {\it entropy function} $F(\vec{q},\vec{e},\vec{v},\vec{u})$ is then defined
as the Legendre transform of $f$ with respect to the charges $e_I$, i.e.
 \bea
F(\vec{q},\vec{e},\vec{v},\vec{u}) &\equiv &
2\pi\, \Big[ e_I q^I- f(\vec{e},\vec{v},\vec{u})
\Big]
\nn\\[1ex]
&=& 2\pi \Big[ e_I q^I
- {v_1 v_2\over 4 G_4} \Big( -{2\over v_1}+{2\over v_2} +\sum_{I=1}^4 \,
{u_I^2 e_I^2\over 2 v_1^2}
+ 4 g^2 \sum_{I<J}^4 \,u_I u_J \Big)\Big]
\;.
\label{defF}
\eea
The near horizon geometry can be found by
extremizing the {entropy function} $F(\vec{q},\vec{e},\vec{v},\vec{u})$
wity respect to $\vec{e},\vec{v}$, and $\vec{u}$:
\bea
{\partial F\over \partial v_{a}}~=~
{\partial F\over \partial u_{I}}~=~{\partial F\over \partial e_I}~=~0 \;.
\label{mineq}
\eea
The first two equations ensure that the metric and the scalar field
equations of motion are satisfied,
 while the last equation defines the black hole electric charges~$q^I$
\be
 q^I~=~
 \frac{\delta}{\delta e_{I}}\,
f(\vec{e},\vec{v},\vec{u})~=~
 {v_2\over 4\,G_4\, v_1 }\, u_I^2 e_I=-{1\over 16 \pi G_4}\int_{S^2} X_I^2\, *F^I
\;.
\label{elec4}
\ee
 In the following we will take $G_4=\ft18$ in such a way that the charges
 $q^{I}$ are normalized to be integers. This normalization is determined
 in Appendix \ref{units} by matching the physical charge units here
 with those coming from string theory brane setups.
 The $G_4$ dependence can be restored by the rescaling (\ref{restoreg4})
 of the physical charges $q^I$.

Evaluating  the entropy function $F$ at the extremum
$(\vec{e}_0(\vec{q}),\vec{v}_0(\vec{q}),\vec{u}_0(\vec{q}))$
one finds the entropy  of the corresponding black hole solution as a
function of the electric charges $\vec{q}$:
\be
S_{\rm BH}(\vec{q})=F(\vec{q},\vec{e}_0(\vec{q}),\vec{v}_0(\vec{q}),\vec{u}_0(\vec{q}))
\;.
\ee
 In practice, the relations (\ref{elec4}) are highly nonlinear and generically hard to invert,
  therefore we will often choose to give an implicit parametrization of the black hole solution,
  its entropy, and the electric charges $q^I$ in terms of $u_{1,2,3}$ and $v_2$
  rather than expressing the entropy directly in terms of the four physical charges~$q^I$.

 It is important to stress that the entropy function formalism applies to
 (in general non-supersymmetric) higher derivative Lagrangians that depend
only on the Riemann and the stress energy tensor but not on their covariant derivatives.
 In this section we consider Einstein gravity, while higher derivative corrections
 to black hole entropies will be considered in section \ref{higherd}.

\subsection{The solution }

As we mentioned in our preliminary discussion,
it is often easier to solve equations (\ref{mineq}), \Ref{elec4} implicitly
in terms of a set of independent parameters
rather than in terms of the four charges $q^{I}$.
We choose parameters $\mu_{I}$
to parametrize the fixed value scalars $u_{1,2,3}$ and the sphere volume $v_2$:
\bea
u_{I}&=&\frac{\mu_{I}}{(\mu_{1}\mu_{2}\mu_{3}\mu_{4})^{1/4}}\;,
\qquad
v_{2}~=~ \ft14\sqrt{\mu_{1}\mu_{2}\mu_{3}\mu_{4}}\;,\label{genq0}
\eea
Plugging (\ref{genq0}) into (\ref{mineq}) and solving for the
remaining variables, one finds the general solution:
\bea
v_{1}&=&
\frac{\ft14\sqrt{\mu_{1}\mu_{2}\mu_{3}\mu_{4}}}
{1+ g^{2}\,\sum_{J<K}\,\mu_{J}\mu_{K}}\;,
\qquad
e_{I} ~=~ \frac{\sqrt{\mu_{1}\mu_{2}\mu_{3}\mu_{4}\,
(1+ g^{2}\,{\textstyle \sum_{I\not=J<K\not=I}}\,\mu_{J}\mu_{K})}}
{2\mu_{I}\,(
1+ g^{2}\,\sum_{J<K}\,\mu_{J}\mu_{K})}
\;,
\nn\\[2ex]
q^{I} &=& \mu_{I}\,\sqrt{1+ g^{2}\,{\textstyle \sum_{I\not=J<K\not=I}}\,\mu_{J}\mu_{K}}
\;.
\label{genq}
\eea
It is easy to check that the equations of motion (\ref{eom4}) are satisfied by
(\ref{ansatz0}), (\ref{genq0}), (\ref{genq}).
Plugging this into the entropy function~\Ref{defF} yields for the black hole entropy
\bea
 S_{\rm BH}(q)
 &=& 2\pi\, \sqrt{\mu_1 \mu_2 \mu_3 \mu_4}
~=~  {\pi v_2 \over G_4} ~=~{1\over 4 G_4} {\cal A}_{\rm hor}
\;,
\eea
in agreement with the Bekenstein-Hawking formula.

In order to express the entropy directly in terms of the electric charges $q^{I}$,
one has to invert the last equation in~\Ref{genq}.
In lowest orders of the gauge coupling this gives rise to the expansion
\be
\mu_I = q^I \left(1-\ft12 g^2 \,\partial_{I}\beta_{3}
+\ft18 g^{4}\,\Big(
\partial_{I}\,(3\beta_{2}\beta_{3}+\beta_{1}\beta_{4})
-2q^I \,\partial^{2}_{I}\,(\beta_{2}\beta_{3})
+4\beta_{4}
 \Big)
+\dots \right)
\;,
\ee
 in terms of the symmetric polynomials
\bea
\beta_{1}={\textstyle\sum_{I}} \,q^{I}\;,\quad
\beta_{2}={\textstyle\sum_{I<J}} \,q^{I}q^{J}\;,\quad
\beta_{3}={\textstyle\sum_{I<J<K}}\, q^{I}q^{J}q^{K}\;,\quad
\beta_{4}=q_{1}q_{2}q_{3}q_{4}\;,
\nonumber
\eea
and with $\partial_{I}=\frac{\partial}{\partial q^{I}}$.
For the entropy this leads to the expansion
\bea
S_{{\rm BH}}
&=& 2\pi\, \sqrt{\beta_4}\;
\Big(
1
- \ft12\,g^2\,\beta_{2}
+\ft18\,g^4\,(3\beta_{2}^{2}+2\beta_{1}\beta_{3}+4\beta_{4})
\label{entropyexp4}
\\[.5ex]
&&\qquad\qquad{}
-\ft{1}{16}\, g^6\,(5\beta_{2}^{3}+9\beta_{1}\beta_{2}\beta_{3}+\beta_{3}^{2}+5\beta_{1}^{2}\beta_{4}+20\beta_{2}\beta_{4})
+\dots\Big)\;.
\nonumber
\eea
The expansion drastically simplifies in two particular cases:

\subsubsection*{Ungauged theory}

At $g=0$ one finds $q^I=\mu_I$ leading to:
\bea
v_1= v_2=\ft14\sqrt{q_1 q_2 q_3 q_4}\;,\quad\;\;
u_{I}=\frac{q^{I}}{(q_1 q_2 q_3 q_4)^{1/4}}\;,\quad\;\;
e_I= \frac{1}{2q^{I}}\, { \sqrt{q_1 q_2 q_3 q_4}}\;,\label{sol4}
\eea
and one recovers the known result
\bea
S_{\rm BH}(\vec{q})&=& 2 \pi\,  \sqrt{q_1 q_2 q_3 q_4}
\;,
\eea
for the entropy in terms of the physical charges.

 \subsubsection*{Equal charges $q^I=q$}

In the case of equal charges, the last equation in \Ref{genq}
can be explicitly solved for $\mu$ and one obtains
the explicit solution
\bea
v_{1}&\!=\!&\frac{\sqrt{1+12 g^2 q^2 }-1}{24g^{2}\sqrt{1+12 g^2 q^2 }}\;,\qquad
v_{2}=\frac{\sqrt{1+12 g^2 q^2 }-1}{24g^{2}}\;,\qquad
u_I=1\;,
\nonumber\\[1ex]
e_{I}&\!=\!&{ q \over 2\sqrt{1+12 g^2 q^2 }} \;,
\eea
and the black hole entropy
\bea
 S_{\rm BH}(q)
 &=& { \pi\, (\sqrt{1+12\, q^2\, g^2\, }-1)\over 3\, g^2 }
\;,
\eea
expressed directly in terms of the electric charges $q$.


\section{AdS$_5$ static black holes}
\label{nonrot5}

Next we consider the $U(1)^3$ gauged supergravity in $d=5$ dimensions.
This theory can be obtained as a truncation of the maximal $N=8$, $SO(6)$ gauged supergravity~\cite{Gunaydin:1985cu}
down to the $U(1)^3$ Cartan subgroup of $SO(6)$. The bosonic action can be written
as
\bea
S &=& {1\over 16 \pi G_5}\int d^5 x \sqrt{-g}\,
\Big[ R-\ft14 X^2_{I} F_{\mu\nu}^I F^{\mu\nu I}
-\ft12\, X_I^{-2}\,\partial_{\mu} X_I \,\partial^{\mu} X_I - V \nn\\
&&\qquad\qquad\qquad\qquad\qquad\qquad{}+
\ft{1}{24}\omega^{\mu\nu\sigma\rho\lambda}\,|\epsilon_{IJK}|\,
F_{\mu\nu}^I\,F_{\sigma\rho}^J\, A^K_\lambda \Big]
\;,
\eea
 with $I=1,2,3$, $\omega^{tr\psi\theta\phi}=-(\sqrt{-g})^{-1}$, and
\be
F_{\mu\nu}^I=2\partial_{[\mu}A^{I}_{\nu]}\;,\qquad
V= -4\, g^2\, \sum_{I=1}^3  X_I \;,\qquad
X_1 X_2 X_3 =1\;.
\ee
The equations of motion derived from this Lagrangian are:
\bea
&& R_{\mu \nu}-\ft12\, X_I^2\, F^I_{\mu \sigma}\, F_{\nu}^I {}^{\sigma}
-\ft12\, X_I^{-2}\,\partial_{\mu} X_I \,\partial_{\nu} X_I
-\ft12\, g_{\mu\nu} \left(R-\ft14 X_I^2 F^{I2}
-\ft12\, (X_I^{-1}\partial X_I)^2
-V \right)=0\;,\nn\\[1ex]
&& {\delta\over \delta X_I} (\ft14 X_I^2 F^{I2}+\ft12\, (X_I^{-1}\partial X_I)^2+V)=0\;,\nn\\[.5ex]
&& \frac{1}{\sqrt{-g}}\,\partial_\mu(\sqrt{-g}\, X_{I}^2 \, F^{\mu \lambda I})+\ft18  \,|\epsilon_{IJK}|\,
\omega^{\mu\nu\sigma\rho\lambda}\,
 F_{\mu\nu}^J\,F_{\sigma\rho}^K =0
\;.\label{eom5}
\eea

We search for non-rotating black holes with near horizon $AdS_2\times S^3$ geometries
\bea
ds^2 &=& v_1 \Big(- r^2\,dt^2+ {dr^2\over r^2}\Big) +v_2 d\Omega_3
\;,\nn\\[1ex]
 X_{I} &=& u_I \;,\qquad A^I ~=~ -e_I \, r\, dt \;, \qquad
 F^I_{0r} = e_I \;, \label{ansatz5}\\[1.2ex]
d\Omega_3 &=& \ft14 \left[d\theta^2+d\psi^2
+ d\phi^2+2 d\phi\, d\psi\, \cos\theta\right]\;,
\;\;  0\leq \psi\leq 2\pi \;,
\;\; 0\leq \phi \leq 4 \pi \;,
\;\; 0\leq \theta \leq \pi\;,
\nn
\eea
 with constants $u_I, e_I, v_a$, and $u_3=1/(u_1 u_2)$.

As before we denote by $f(\vec{e},\vec{v},\vec{u})$, the
supergravity action evaluated on the background (\ref{ansatz5})
and integrated over the three-sphere:
\be
f(\vec{e},\vec{v},\vec{u}) \equiv
\int d\theta d\phi\,d\psi \,\sqrt{-g}\, {\cal L}(\vec{e},\vec{v},\vec{u})
\;.
\ee
The {\it entropy function} $F(\vec{q},\vec{e},\vec{v},\vec{u})$ is again defined
as the Legendre transform of $f$ with respect to the charges $e_I$, i.e.
\bea
F(\vec{q},\vec{e},\vec{v},\vec{u}) &\equiv &
2\pi\, \Big[ e_I q^I-f(\vec{e},\vec{v},\vec{u})
\Big]
\label{defF5}\nn\\[1ex]
&=& 2\pi \Big[ e_I q^I
- {\pi\over 8 G_5} v_1 v_2^{3\over 2} \Big( -{2\over v_1}+{6\over v_2} +\sum_I \,{u_I^2 e_I^2\over 2 v_1^2}
+ 4 g^2 \sum_I \,u_I \Big)\Big]
 \;.
 \eea
Note that the Chern-Simons term does not contribute to the action
in the near horizon geometry~(\ref{ansatz5}).
The near horizon geometry is again found by extremizing $F$:
\bea
{\partial F\over \partial e_I}~=~{\partial F\over \partial v_{a}}~=~{\partial F\over \partial u_{I}}~=~0
\;.
\label{mineq5}
\eea
The first equation defines the electric charges $q^{I}$ as
\be
 q^I~=~
 \frac{\delta}{\delta e_{I}}\,
f(\vec{e},\vec{v},\vec{u})~=~{ \pi\,v_2^{3\over 2}\over 8\, G_5\, v_1} u_I^2 e_I
=-{1\over 16 \pi G_5}\int_{S^3} X_I^2\, *F^I
\;.
\label{elec5}
\ee
In the following we will take $G_5={\pi\over 4}$ in such a way that the charges
 $q^{I}$ are normalized to be integers. This normalization is justified
 in Appendix \ref{units}.
 The $G_5$ dependence can be restored by the rescaling (\ref{restoreg5})
 of the physical charges $q^I$.

Evaluating the entropy function at the minimum $(\vec{e}_0(\vec{q}),\vec{v}_0(\vec{q}),\vec{u}_0(\vec{q}))$
one finds the entropy  of the corresponding black hole solution as a function of the electric charges
$\vec{q}$.
\be
S_{BH}(\vec{q})=F(\vec{q},\vec{e}_0(\vec{q}),\vec{v}_0(\vec{q}),\vec{u}_0(\vec{q}))
\;.
\ee

\subsection{ The solution }

In analogy to the four-dimensional case above we introduce three independent parameters $\mu_I$
to parametrize $u_{1,2}$ and $v_2$. The general solution of~\Ref{mineq5} can then be written as
\bea
u_{I}&\!=\!&\frac{\mu_{I}}{(\mu_{1}\mu_{2}\mu_{3})^{1/3}}\;,\qquad
v_{2}= (\mu_{1}\mu_{2}\mu_{3})^{1/3}\;,
\nonumber\\[1ex]
v_{1}&\!=\!& \frac{(\mu_{1}\mu_{2}\mu_{3})^{1/3}}{4(1+ g^{2}\,\sum_{J}\mu_{J})}\;,\quad
e_{I}= \frac{\sqrt{\mu_{1}\mu_{2}\mu_{3}
(1+ g^{2}\,\sum_{J\not=I}\mu_{J})}}{2\mu_{I}(1+ g^{2}\,\sum_{J}\mu_{J})}
\;,\qquad
\nonumber\\[2ex]
q^{I}&=&  \mu_{I}\,\sqrt{1+ g^{2}\,{\textstyle \sum_{J\not=I}}\mu_{J}}
\;.
\label{sol5genm}
\eea
It is easy to check that equations of motion (\ref{eom5}) are satisfied by
(\ref{ansatz5}), (\ref{sol5genm}).
With this solution we obtain from~\Ref{defF5}
for the entropy of the black hole
\bea
 S_{\rm BH} &=&
 2 \pi \sqrt{\mu_1 \mu_2 \mu_3 }
 ~=~{\pi^2 v_2^{3\over 2}\over 2 G_5}~=~{1\over 4 G_5} {\cal A}_{\rm hor}
\;,
\eea
again in agreement with the Bekenstein-Hawking formula.

In lowest order of the gauge coupling we obtain the following expansion
\be
\mu_I = q^I\, \Big(
1-\ft12\, g^2\,\partial_{I}\beta_{2}
+\ft1{8}\,g^{4}\,(\partial_{I}(3\beta_{1}\beta_{2}+5\beta_{3})-4\beta_{2})
+\dots \Big)
\;,
\ee
in terms of the symmetric polynomials
\bea
\beta_{1}=\sum_{I} q^{I} \;,\quad
\beta_{2}=\sum_{I<J} \,q^{I}q^{J}\;,\quad
\beta_{3}=q_{1}q_{2}q_{3}\;.
\nonumber
\eea
For the entropy this implies
\bea
S_{{\rm BH}}
&=&
 2 \pi\sqrt{\beta_3 }\,
\Big(
1
- \ft12\,g^{2}\,\beta_{1}
+\ft18\,g^4\,(3\beta_{1}^{2}+2\beta_{2})
-\ft{1}{16}\,g^6\,(5\beta_{1}^{3}+9\beta_{1}\beta_{2}+5\beta_{3})
\nonumber\\[.5ex]
&&\qquad\qquad\qquad{}
+\ft{1}{128}\,g^8\,(35\,\beta_{1}^{4}+116\beta_{1}^{2}\beta_{2}+20\beta_{2}^{2}+136\beta_{1}\beta_{3})
+\dots\Big)\;.
\label{entropyexp5}
\eea
Again drastic simplifications occur for
$g=0$ and for all charges equal $q^I=q$:

\subsubsection*{Ungauged theory}

At $g=0$ we have $\mu_{I}=q^{I}$ and the solution takes the explicit form
\bea
v_2=4 v_1=(q_1 q_2 q_3)^{1\over 3}\;,
\qquad
 u_I=\frac{q^{I}}{(q_1 q_2 q_3)^{1\over 3}}\;,
 \qquad
 e_I={1\over 2 q^I}\, \sqrt{ q_1 q_2 q_3}
 \;,
\label{sol5ung}
\eea
and the black hole entropy is simply given as
\bea
 S_{BH}&=&2\pi \sqrt{ q_1 q_2 q_3}
\;.
\eea

\subsubsection*{Equal charges $q^I=q$}

In this case the above formulas reduce to
\bea
v_{1}&\!=\!& \frac{\mu}{4(1+3g^{2}\,\mu)}\;,\qquad
v_{2}=\mu\;,\qquad
u_{I}=1\;,\nonumber\\[1ex]
e_{I}&\!=\!& \frac{\sqrt{\mu+2g^{2}\,\mu^2}}{2\,(1+3g^{2}\,\mu)}
\;,\qquad
q=~ \mu\,\sqrt{1+2g^{2}\,\mu}
\;,
\label{sol5m}
\eea
with black hole entropy
\bea
 S_{\rm BH}
&=& 2\pi\,\mu^{3/2}
~=~{\pi^2 v_2^{3\over 2}\over 2 G_5}~=~{1\over 4 G_5} {\cal A}_{\rm hor}\;,
\eea
expressed in terms of a single parameter $\mu$.
If instead we choose to express $S_{\rm BH}$ directly in terms of the charges $q$
we have to invert the last equation in~\Ref{sol5m}. A closed form
for the entropy in this case is given by the more involved expression
\bea
S_{\rm BH}&=&
\frac{2 \,\sqrt{3}\,q^{3/2}}
{\sqrt{\sin\phi+\sqrt{3}\cos\phi+(2/3)\,\sin3\phi }}
\;,\qquad
\phi=\ft13\,{\rm arcsin}(3\sqrt{3}\, q\,g^{2})
\;.
\eea

\section{Rotating black holes in AdS$_5$}
\label{rot5}

 Finally we consider rotating black holes with squashed $AdS_2\times S^3$
near horizon geometry\footnote{Squashing here refers to the full product,
still the metric has the AdS$_{2}$ isometries, see~\cite{Davis:2005ys}.}
\bea
ds^2 &=& v_1\, \Big( -r\,dt^2+{dr^2\over r^2}\Big) +
\ft14 v_2 \left[\sigma_1^2+\sigma_2^2
+ v_3(\sigma_3  -\alpha \,r\, dt)^2\right]\;,
\nn\\
 X_{I} &=& u_I \;,    \nn\\[1ex]
A^I &=& -e_I \, r\, dt+p_I \, \sigma_3\;,
\qquad F^I_{0r}=e_I\;, \quad\quad
F^I_{\psi\theta}=p_I \, \sin\theta\;,\nn\\[1ex]
\sigma_1^2+\sigma_2^2&=& d\theta^2+\sin^2 \theta d\psi^{2}\;,
\qquad \sigma_3=d\phi+\cos\theta d\psi\;,
\label{ansatzr5}
\eea
 with constants $u_I, e_I, p_I, v_a$, and $\alpha$. The constants $\alpha,v_3$ and $p_I$
parametrize the breaking from the $SO(4)$ isometry of the non-rotating
solution down to $SU(2)\times U(1)$ once the angular momentum is turned on.

The entropy function is then given by \cite{Astefanesei:2006dd}
 \bea
 F(\vec{q},\vec{e},\vec{v},\vec{u}) &\equiv&
2\pi\, \left( \alpha J+ e_I\,\hat{q}^I
- f(e_I,\alpha,v_a,u_I)
\right)\nn\\[2ex]
&=& 2\pi \Big[\alpha J+e_I \hat{q}^I+{\pi\over 3 G_5}\, |\epsilon_{IJK}| e_I p_J p_K
\label{defF5r}\\
&&
\!\!\!\!\!\!\!\!\!\!\!\!\!\!\!
- {\pi v_1 v_2^{3\over 2} v_3^{1\over 2}\over 8 G_5}\Big( -{2\over v_1}+{8-2 v_3\over v_2}+
{v_2 v_3 \alpha^2\over 8 v_1^2} +
\sum_I {e_I^2 u_I^2\over 2 v_1^2}
-8 \sum_I {p_I^2 u_I^2\over v_2^2}
+4 g^2 \sum_I u_I \Big)\Big]\;.
\nn
 \eea
Notice that now also the Chern-Simons term contributes to the action.
The fact that the Chern-Simons term depends explicitly on the potential $A_\mu$
rather than on the field strength $F_{\mu\nu}$
requires a slight modification of Sen's algorithm.
First, the presence of the Chern-Simons term modifies the definition of the electric
charge $q^I$. This can be easily implemented in the entropy function
by the redefinition $q^I=\hat{q}^I+c^I$ with $c^I$ chosen such that $q^I$ are conserved
quantities. Luckily the $c^I$ induced by the Chern-Simons term are independent
of $e_I$, $u_I$, and $v_a$ such that this modification will modify neither the
attractor equations nor the black hole entropy.
Second, due to the presence of the Chern-Simons term, the equations of
motion for $A_\phi^{I}$
\bea
0&=&\frac{1}{\sqrt{-g}}
 \partial_\mu(\sqrt{-g}\, X_{I}^2 \, F^{\mu \phi I})+\ft18 \,|\epsilon_{IJK}|\,
\omega^{\mu\lambda\sigma\rho\phi}\,
 F_{\mu\lambda}^J\,F_{\sigma\rho}^K
\;,\nn\\
&=& {\alpha\,u_I^2\, e_I\over v_1^2} -{16 \,u_I^2\,p_I\over v_2^2}
- {8\over \,v_1 \, v_2^{3\over 2}\, v_3^{1\over 2}} \, |\epsilon_{IJK}|\,
 e^J\, p^K \;,
 \label{eqp}
\eea
are no longer automatically satisfied as a mere consequence of the extremization equations
\bea
{\partial F\over \partial \alpha}={\partial F\over \partial e_I}
={\partial F\over \partial u_I}={\partial F\over \partial v_a}=0
\;.
\label{mineq5r}
\eea
Rather, equations \Ref{eqp} have to be considered in addition to
the extremization equations \Ref{mineq5r}
and determine the fluxes $p_I$ in the ansatz (\ref{ansatzr5}).

The resulting solution describes the near horizon geometry of a black hole
with electric charges $q^I$ and angular momenta $J$ given by
\bea
 q^I &=&
 \frac{\delta}{\delta e_{I}}\,
f(\vec{e},\vec{v},\vec{u})-{\pi\over 6\, G_5}\, |\epsilon_{IJK}|\, p_J p_K~=~
{\pi\, v_2^{3\over 2}\,v_3^{1\over 2}\, u_I^2\over 8 \,G_5 \,v_1} \,e_I
-{\pi\over 2 \,G_5}\, |\epsilon_{IJK}|\, p_J p_K\nn\\[.5ex]
&=&-{1\over 16 \pi G_5}\int_{S^3} (X_I^2\, *F^I+\ft12 |\epsilon_{IJK}| \, F^J\wedge A^K)
\;, \nn\\[3ex]
  J &=&
 \frac{\delta}{\delta \alpha}\,
f(\vec{e},\vec{v},\vec{u})~=~{\pi \,v_2^{5\over 2}\, v_3^{3\over 2}
\over 32 \,G_5\, v_1} \alpha
={1\over 16 \pi G_5}\int_{S^3} *dK
\;.
\label{elec5r}
\eea
Here $K={\partial \over \partial \phi}$ denotes the Killing vector associated with the
angular rotation. The shift $c^I=-{\pi\over 6 G_5}\, |\epsilon_{IJK}|\, p_J p_K$ has been
chosen in such a way that the integrand in the definition of $q^I$ is closed
on the mass shell
\be
d(X_I^2\,*F^I)+\ft12  |\epsilon_{IJK}| \, F^J\wedge F^K=0\;.
\ee
This allow us to identify $q^I$  with the conserved charge\footnote{J.F.M. thanks
L.Alvarez-Gaume and C.N. Pope for useful discussions on this point.}.
As we explained before neither the solution nor the entropy depends on the $c^I$'s.
 In the rest of this section we describe the different subcases for which
we can give explicit solutions to (\ref{eqp}), (\ref{mineq5r}).

\subsection{BPS black holes}
\label{ssusy}

Let us first discuss the case of extremal BPS rotating black holes.
These black hole solutions have been found in~\cite{Gutowski:2004yv}.

In this case, we can give the general solution of~\Ref{eqp}, \Ref{mineq5r}
again in terms of three independent parameters $\mu_{I}$
and their symmetric polynomials
\bea
\gamma_{1}=\sum_{I} \mu_{I} \;,\quad
\gamma_{2}=\sum_{I<J} \,\mu_{I}\mu_{J}\;,\quad
\gamma_{3}=\mu_{1}\mu_{2}\mu_{3}\;,
\nonumber
\eea
as follows
\bea
u_{I}&\!=\!&\frac{\mu_{I}}{\gamma_3^{1/3}}\;,\qquad
v_{1}\!=\! \frac{\gamma_3^{1/3}}{4(1+ g^{2}\,\gamma_{1})}\;,\quad
v_{2}= \gamma_3^{1/3}\;,\nonumber\\[2ex]
v_{3} &=& 1+g^{2}\,\gamma_{1}-\frac{g^{2}\,\gamma_{2}^{2}}{4\gamma_{3}}\;,\qquad
\alpha =
\frac{g\,\gamma_{2}}
{(1+ g^{2}\,\gamma_{1})
\sqrt{4\gamma_{3}\,(1+g^{2}\,\gamma_{1})
-g^{2}\,\gamma_{2}^{2}}}
\;,
\nonumber\\[2ex]
e_{I}&\!=\!&  \frac{
\sqrt{4\gamma_{3}\,(1+g^{2}\,\gamma_{1})
-g^{2}\,\gamma_{2}^{2}}}
{4\mu_{I}\,(1+ g^{2}\,\gamma_{1})}
\;,\qquad
p_{I}=
\ft14\,g\,(\gamma_{1}-\mu_{I})-\frac{g\,\gamma_{3}}{4\mu_{I}^{2}}\;,
\nonumber\\[2ex]
q^{I}&=&
\mu_{I}+\ft12\,g^2 \,  \mu_I (\gamma_1-\mu_I)-  {g^2\,\gamma_3\over 2\mu_I}
\quad\quad J=
\frac{g\,\gamma_{2}\,
(4\gamma_{3}\,(1+g^{2}\,\gamma_{1})
-g^{2}\,\gamma_{2}^{2})}{16 \gamma_{3}}
\;
\label{sol5genrotm}
\eea
Plugging this into \Ref{defF5r} we obtain for the entropy
\bea
 S_{\rm BH}~=~
 2\pi\, \sqrt{\gamma_{3}\,(1+g^{2}\,\gamma_{1}) -\ft14\,g^{2}\,\gamma_{2}^{2}} ~=~
 {\pi^2 v_2^{3\over 2} v_3^{1\over 2}\over 2 G_5}~=~
 {1\over 4 G_5} {\cal A}_{\rm hor}
 \;,
\label{solrotBPS}
\eea
reproducing the result of~\cite{Gutowski:2004yv}.
In order to compare the results it is helpful to note that the
parametrization of the squashed $AdS_2\times S^3$ near horizon geometry
given in~\cite{Gutowski:2004yv}
\bea
ds^2_{\rm BPS} &=&-f^2\, dT^2+2 f^2 \,w\, dT \sigma_3+f^{-1} \,b^{-1}\, dR^2+
\ft14\, R^2\, f^{-1}\,(\sigma_1^2+\sigma_2^2+c \, \sigma_3^2)\;,
\nonumber\\[1ex]
f &=& R^2 \gamma_3^{-{1\over 3}}\;,
\quad
w=-{\gamma_2 g\over 4 R^2}\;,
\quad
b=1+g^2 \gamma_1\;, \quad
c=1+g^2 \gamma_1-{g^2 \gamma_2^2\over 4 \gamma_3}
\;,
\label{gutsolA}
\eea
translates into the standard form (\ref{ansatzr5}) with
\bea
 r =R^2\;, \quad
 dt = {2\,b\over \sqrt{\gamma_3\, c}}\, dT\;,
 \quad
 v_1 ={\gamma_3^{1\over 3}\over 4 b}\;,\quad
 v_2=\gamma_3^{-{1\over 3}}\;,\quad
 v_3=c\;,\quad
 \alpha ={\gamma_2\, g\over 2\, b \sqrt{\gamma_3 \, c}}
 \;.
 \nn
 \eea
in agreement with (\ref{sol5genrotm}).

\subsection{Non-extremal black holes}

These considerations can be extended to non-extremal black holes.
For simplicity we focus on the case of equal charges $q^{I}=q$.
The general solution of equations~\Ref{eqp}, \Ref{mineq5r}
can then be expressed in terms of two independent parameters $\mu$, $\omega>1$
as
\bea
 u_I &\!=\!& 1\;,\qquad
 v_{1}= \frac{\mu}{4(1+3g^{2}\mu)}\;,\quad
v_{2}=\mu\;,\quad
v_{3} = \mu^{-3} \Delta_s^2\;,
\label{nearot}\\[1ex]
e_{I}&\!=\!&
\frac{\Delta_s}{2\mu (\w-1)(1+3g^{2}\mu)}\;,
\qquad
p_{I} =
\frac{\Delta_\alpha}{2\mu (1+w)}\;,
\qquad
\alpha =
\frac{\Delta_\alpha}{\Delta_s(1+3g^{2}\mu)}
 \;,
\nn\\[1ex]
J &=&
 \ft12 \, \mu^{-3}\, \Delta_\alpha \Delta_s^2
\;, \qquad
q^{I} = {2\mu\over \w}+2\, g^2\, \mu^2 {(\w-1)\over \w^2}
\;,
\nonumber
 \eea
with
\bea
\Delta_\alpha &=&{\mu(\omega+1)\over \omega^2}\, \sqrt{2\mu \omega(\omega-2)
+4 g^2 \mu^2(\w^2-2 \w+1)}\;,\nn\\
\Delta_s &=&{\mu(\omega-1)\over \omega^2}\, \sqrt{2\mu \omega(\omega+2)
+2 g^2 \mu^2 (\w^2+2 \w-2)}\;.
\eea

Plugging this into \Ref{defF5r}, we find for the entropy
\bea
 S_{\rm BH}=2\pi\Delta_s
 ~=~
 {\pi^2 v_2^{3\over 2} v_3^{1\over 2}\over 2 G_5}=
 {1\over 4 G_5} {\cal A}_{\rm hor}
 \;.
\label{solrot}
\eea

It is interesting to note that although for a generic choice of the
parameters $\mu$, $\omega$ the
black hole solution found here is non-supersymmetric, the charges
can be chosen in such a way that the BPS bound is saturated.
More precisely, for the particular value $\omega=2$
the formulas obtained here reduce to
\bea
u_{I}&\!=\!&1\;,\quad
e_{I}=  \frac{
\sqrt{\mu}\sqrt{4+3g^{2}\,\mu}}
{4\,(1+ 3g^{2}\,\mu)}
\;,\quad
p_{I}=
\ft14\,g\,\mu\;,
\quad
q^{I}= \mu+\ft12\,g^{2}\,\mu^{2}\;,
\nonumber\\[2ex]
v_{1}&\!=\!& \frac{\mu}{4(1+ 3g^{2}\,\mu)}\;,\quad
v_{2}= \mu\;,\quad
v_{3}=1+\ft34\,g^{2}\,\mu\;,\quad
\alpha =
\frac{3g\,\sqrt{\mu}}
{(1+ 3g^{2}\,\mu)
\sqrt{4+3g^{2}\mu)}}
\;,
\nonumber\\[2ex]
J&=&
\frac3{16}\,g\,\mu^{2}\,(4+3g^{2}\mu)
\;,
\qquad
S_{\rm BH}= 2\pi\,\mu^{3/2}\,\sqrt{1+\ft34\,g^{2}\,\mu}
\;,
\label{sol5equal}
\eea
that agrees with the general BPS solution \Ref{sol5genrotm} after taking all charges
equal $\mu_i=\mu$.

 Another interesting limit of the solution~\Ref{nearot} is the unrotating case
studied in last section. This is given by setting
\bea
\omega=1+\frac{1}{\sqrt{1+2g^{2}\mu}}\;.
\eea
Indeed it is straightforward to check that
at this value the above formulas reduce to~\Ref{sol5m}.

\section{Higher derivative terms}
\label{higherd}

Finally we consider asymptotically Anti-de Sitter black hole horizons in
higher derivative gravity. In contrast to the case of Poincar\'e supergravities,
higher derivative couplings in gauged supergravities were rarely studied in
the string literature. Awaiting more realistic Lagrangians here we
illustrate the entropy formalism in an archetype toy
example: the Einstein-Maxwell system in presence of a Gauss-Bonnet term
and a cosmological constant
\be
S={1\over 16 \pi G_d} \int d^d x \sqrt{-g} \left( R-\ft14 F^2+\Lambda
+a \,{\cal L}_{\rm GB}
\right)
\;,
\label{actgb}
\ee
 with the Gauss-Bonnet term
\be
{\cal L}_{GB}=R_{\mu\nu\sigma\rho}R^{\mu\nu\sigma\rho}-4 R_{\mu\nu}R^{\mu\nu}+R^2
\;.
\ee
 The parameter $a$ measures the deviation from Einstein gravity and it depends on the
particular string model under consideration.\footnote{See~\cite{o1,o2,o3} for an analysis of the
boundary terms needed by the regularization of the action for Einstein-Gauss-Bonnet-AdS
gravity.
}

  The equations of motion following from (\ref{actgb}) are:
\bea
 && R_{\mu\nu}-\ft12 F_{\mu \sigma} F_{\nu}{}^\sigma+
 a\,{\delta {\cal L}_{\rm GB}\over \delta g^{\mu\nu}}-\ft12 g_{\mu\nu}
 \left( R-\ft14 F^2+\Lambda+a\, {\cal L}_{\rm GB} \right)=0\;,
 \nn\\[1ex]
 && \partial_\mu (\sqrt{-g} F^{\mu\nu} )=0 \label{eomd}
 \;,
 \eea
with
\be
{\delta {\cal L}_{\rm GB}\over \delta g^{\mu\nu}}~=~2(R_{\mu\sigma\rho\delta}
\, R_{\nu}{}^{\sigma\rho\delta}-2\, R^{\rho\sigma}\,R_{\mu\rho\nu\sigma}-
2\, R_{\mu}^\sigma\, R_{\nu \sigma}+R\, R_{\mu\nu})
\;,
\ee
up to total derivatives.
We look for $AdS_2\times S^{d-2}$ near horizon geometries:
 \bea
ds^2 &=& v_1 \left(- r^2\,dt^2+ {dr^2\over r^2}\right) +
v_2\, d\Omega_{d-2} \;,\qquad
 F_{0r} ~=~ e\;.
 \eea
The extremization equations of the entropy function can now be explicitly solved
in the different space-time dimensions.

\subsubsection*{d=4}

In four dimensions, evaluating the entropy function for this system yields
\bea
F(q,e,\vec{v}) &\equiv &
2\pi\, \left[ e q- {v_1 v_2\over 4 G_4} \,
 \left( -{2\over v_1}+{2\over v_2}-{8 a\over v_1 v_2}
+ { e^2\over 2 v_1^2}+ \Lambda \right)\right]
\;.
\label{f4}
\eea
The extremum of $F(q,e,\vec{v})$ (for a fixed $q$) can be conveniently parametrized
in terms of  $v_2$:
 \bea
 v_1 &=& {v_2\over 1+v_2 \Lambda}\;, \qquad
 e~=~\sqrt{ 2 v_2(2+v_2 \Lambda)\over (1+v_2 \Lambda)}\;,\qquad
q~=~{1\over 2 G_4}\sqrt{v_2(1+\ft12 v_2 \Lambda)} \;.
 \label{extrem4}
\eea
Plugging (\ref{extrem4}) into the entropy function (\ref{f4}) one finds
the black hole entropy
 \be
S_{\rm BH}~=~{\pi \over G_4}\,(v_2+4 a)\;.
 \ee
  The $a$-term gives the deviation from the area law due to the
Gauss-Bonnet term. Interestingly, the presence of the
Gauss-Bonnet term in $d=4$ does not modify the near horizon solution but only
the black hole entropy. This is consistent with the fact that
in $d=4$ the $a$-dependent term in the equations of motion (\ref{eomd}) cancels
once evaluated on $AdS_2\times S^2$.
In $d=5$ this will be different as we shall see.

\subsubsection*{d=5}

In five dimensions the entropy function is given by
\bea
F(q,e,\vec{v}) &\equiv &
2\pi\, \left[ e q- {\pi v_1 v_2^{3\over 2}\over 8 G_5} \,
 \left( -{2\over v_1}+{6\over v_2}-{24 a\over v_1 v_2} + { e^2\over 2 v_1^2}+ \Lambda \right)\right]
 \;.
 \label{f5}
 \eea
The extremum of $F(q,e,\vec{v})$ (for a fixed $q$) can be conveniently parametrized
in terms of the sphere radius $v_2$:
 \bea
 v_1 = {v_2+4 a\over 4-v_2 \Lambda}\;,
 \quad
 e=\left({v_2+4 a\over 4-v_2 \Lambda}\right)\sqrt{ 12\, v_2^{-1}-2\,\Lambda}\;,
\quad
q={\pi v_2\over 4 G_5}\sqrt{3-\ft12 v_2 \Lambda } \;.
\label{extrem}
\eea
Plugging (\ref{extrem}) into the entropy function (\ref{f5}) one finds the black hole
entropy
 \be
S_{\rm BH}~=~{\pi^2 v_2^{1\over 2}\over 2 G_5}\,(v_2+12 a)
\;.
 \ee
The $a$-dependent term represents the deviation from the area law due to the Gauss-Bonnet term.

\section{Conclusions}

In this paper we applied the entropy formalism to the case
of gauged supergravities which admit asymptotically AdS
electrically charged black holes with $AdS_2\times S^{d-2}$ horizons.
Using Sen's algorithm we have determined the fixed near-horizon
geometries for four and five-dimensional static black holes,
for rotating five-dimensional black holes and finally
for AdS black holes with higher derivative corrections of Gauss-Bonnet
type. In each case we find horizons with fixed scalars, AdS and sphere radii,
determined entirely in terms of the gauge coupling, the
black hole electric charges and the angular
momentum.

The explicit dependence on the gauge potential via
the Chern-Simons term in the five-dimensional gauged supergravity requires a
slight modification of the entropy function algorithm.
We have illustrated this in the case of five-dimensional
rotating black holes. Once the black hole rotates, magnetic fluxes $p_I$ should be
turned on and the Chern-Simons term starts contributing to the action.
The inclusion of this term leads to a redefinition of the black hole electric charge
$q^I\to q^I+c^I$ with $c^I$ depending only on the magnetic fluxes and not
on the metric or scalar fields. This implies in particular that neither the
attractor equations nor the entropy depends on $c^I$ and therefore $c^I$ can be adjusted
to account for the Chern-Simons correction to the electric charge.
The fluxes $p_I$ are determined by an extra constraint coming from the gauge field
equations of motion to be imposed in addition to the extremization conditions of the entropy function.
Nicely, this leads again to a family of black hole near horizon solutions parametrized
only by the black hole electric charges and the angular momentum.

In the case of Einstein gravity, the near horizon geometries derived
here can be recovered by considering the zero temperature
limit of the general black hole solutions
 \cite{Behrndt:1998jd,Cvetic:1999xp,Duff:1999gh,Sabra:1999ux,Gutowski:2004ez,Gutowski:2004yv,Cvetic:2004hs,Cvetic:2004ny,Cvetic:2005zi,Chong:2005da,Chong:2005hr,Kunduri:2006ek}.
In this limit one finds a single horizon with
$AdS_2\times S^{d-2}$ topology. We stress the fact that
in the gauged theory, zero temperature black holes are not
necessarily supersymmetric. A non-BPS black hole solution is known
to be classified by its charges (electric charge, angular momentum, etc.)
and its mass. The condition of zero temperature relates the black
hole mass to its charges. This implies that
 there is a unique black hole solution with  $AdS_2\times S^{d-2}$ horizon for
a given choice of the charges. This is precisely the result coming from
extremizing the entropy function. The precise matching between
our solutions and the $T\to 0$ limit
of the general non-extremal black hole solutions is shown explicitly in appendix
\ref{scomparison} for static black holes
and in section \ref{ssusy} for the five-dimensional BPS case \cite{Gutowski:2004yv}.

It is tempting to speculate about the generalization of the
expressions for the entropy~\Ref{entropyexp4},~\Ref{entropyexp5}
to the full ${\cal N}=8$ theories.
For the ungauged case $g=0$ it is well
known that the first term in the expansions
is replaced by the quartic and cubic invariants of the global symmetry
groups $E_{7}$ and $E_{6}$, respectively~\cite{Kallosh:1996uy}.
The gauging of the theories
is most conveniently
described in terms of an embedding tensor which
parametrizes the deformation in order $g$ and
comes in a particular representation of the
global symmetry groups~\cite{deWit:2002vt}.
This suggests that e.g.~the second term in
the expansion~\Ref{entropyexp4}
will be replaced by an $E_{7}$ invariant
built from six charges and two embedding tensors.
Indeed, there is a single nontrivial $E_{7}$ invariant
combination of these representations which might thus
generalize the expansion~\Ref{entropyexp4} to lowest orders.\\
It would be nice to explore the implications of our results to gauge/gravity
holographic correspondences.
In particular, the $AdS_5$ entropy formula provide explicit predictions
on the partition function of gauge invariant operators in ${\cal N}=4$ SYM.
In addition the $AdS_2\times S^{d-2}$ solutions found here can be used as
starting points of new holographic relations between quantum mechanical systems
living on the $AdS_2$ boundary and the gravity physics near the horizon.
\\
 We hope to come back to some of these issues in the near future.
\bigskip

{\bf Acknowledgement}
We thank L. Andrianopoli, L. Alvarez-Gaume, S. Bellucci, G. Dall'Agata, S. Ferrara, D. Martelli, C.N. Pope
and A. Santambrogio for useful discussions.
This work is supported in part by the European RTN Program MRTN-CT-2004-503369
and the DFG  grant SA 1336/1-1.

\begin{appendix}

\section{Physical charge units}
\label{units}

In this appendix we explain the normalization of physical charges adopted in the text.
Electric charge units do not depend on the coupling constant $g$, therefore we
can restrict ourselves to the ungauged limit $g=0$. The five and four-dimensional
supergravities studied here can be embedded into compactifications of type II
supergravities on $T^5$ and $T^6$ respectively.
The black hole solutions in this limit reduce to the well known 3- and 4-charge black hole solutions
of the maximal supergravities.
Here we normalize our charges in such a way as to match the electric charge units
coming from black holes built out of branes in string theory.
The formulas in this appendix follow the notations and conventions in \cite{Peet:2000hn}.
 We refer the reader to this reference for further details and a complete list of references
 on the subject.

\subsubsection*{Newton constant}
\be
G_d={G_{10}\over (2\pi)^{10-d} V_{10-d}} \;,
\qquad
G_{10}=8\pi^6 g_s^2 \ell_s^8
\;,
\ee
with string length $\ell_s=\sqrt{\alpha'}$, string
coupling constant $g_s$, and the volume $V_{10-d}$ of the compactification
manifold.

\subsubsection*{4-charge black hole}

 The Einstein metric of a 4-charge black hole in $d=4$ dimensions can be written
 as
 \bea
 ds^2 &=&-(H_1 H_2 H_3 H_4)^{-{1\over 2}}\, dT^2+(H_1 H_2 H_3 H_4)^{{1\over 2}}\,
 (dr^2+r^2 d\Omega_2)\;,\nn\\
H_i &=& 1+{c_i N_i\over r}\;,
 \eea
 with integers $N_i$ counting the number of brane constituents and some
constants $c_i$ parametrizing the brane tension. In the near horizon $r\to 0$, the black
hole geometry becomes
\bea
 ds^2 &=&-(c N_1 N_2 N_3 N_4)^{-{1\over 2}}\, r^2\,dT^2+(c N_1 N_2 N_3 N_4)^{{1\over 2}}\,
 {dr^2\over r^2} + (c N_1 N_2 N_3 N_4)^{{1\over 2}} d\Omega_2
 \;,
 \label{bh4}
 \eea
 with
\be
c= c_1 c_2 c_3 c_4={g_s^4 \ell_s^{16} \over 16 V_6^2}= 4 G_4^2
\;.
\ee
Notice that although $c_i$ depends on the type of brane constituent and on the
string model, $c$ is a $U$-duality invariant quantity that depends only on $G_4$.

 After a rescaling of $dT$ the metric (\ref{bh4}) can be put into our
 standard $AdS_2\times S^2$ form (\ref{ansatz0}) with
\be
v_1=v_2=\sqrt{ c N_1 N_2 N_3 N_4}=2 G_4 \sqrt{  N_1 N_2 N_3 N_4}\;.
 \label{sol42}
\ee
 Taking $G_4={1\over 8}$ and comparing (\ref{sol42}) with (\ref{sol4}), one finds
 agreement with the identification $q_i=N_i$, i.e. the $q_i$ are integers.
It is important to note that $G_4$ can be reabsorbed by a simultaneous rescaling
of $q_i$ and $S_{\rm BH}$. Therefore the $G_4$ dependence in the main text
 can be restored by sending
 \be
  q_i \to (8 G_4) \,q_i \;,\quad\quad S_{\rm BH}\to (8 G_4)\, S_{\rm BH}\;.
 \label{restoreg4}
 \ee
Clearly, the $q_i$'s defined in this way will not be integers.

\subsubsection*{\bf 3-charge black hole}

 The Einstein metric of a 3-charge black hole in $d=5$ dimensions can be written
 as
 \bea
 ds^2 &=&-(H_1 H_2 H_3)^{-{2\over 3}}\, dT^2+(H_1 H_2 H_3)^{{1\over 3}}\,
 (dR^2+R^2 d\Omega_3)\;,\nn\\
H_i &=& 1+{c_i N_i\over R^2}\;.
 \eea
 In the near horizon $r=R^2\to 0$, the black hole geometry becomes
\bea
 ds^2 &=&-(c N_1 N_2 N_3)^{-{1\over 2}}\, r^2\,dT^2+(c N_1 N_2 N_3)^{{1\over 2}}\,
 {dr^2\over 4 r^2} + (c N_1 N_2 N_3)^{{1\over 2}} d\Omega_3\;,
 \label{bh5}
 \eea
 with
\be
c= c_1 c_2 c_3 ={g_s^4 \ell_s^{16} \over  V_6^2}= \Big({4 G_5\over \pi}\Big)^2
\;.
\ee
Again $c$ is a $U$-duality invariant quantity depending only on $G_5$.

 After a rescaling of $dT$ the metric (\ref{bh5}) can be put into the
 standard $AdS_2\times S^3$ form (\ref{ansatz5}) with
\be
v_2=4 v_1=(c N_1 N_2 N_3)^{1\over 3}=  \Big({4 G_5\over \pi}\Big)^{2\over 3}
( N_1 N_2 N_3)^{1\over 3}\;. \label{sol52}
\ee
 Taking $G_5={\pi\over 4}$ and comparing (\ref{sol52}) with (\ref{sol5ung}), one finds
 agreement with the identification $q_i=N_i$, i.e. the $q_i$ are integers.

It is important to note that $G_5$ can be reabsorbed by a simultaneous rescaling
of $q_i$ and $S_{\rm BH}$. Therefore the $G_5$ dependence in the main text
 can be restored by sending
 \be
  q_i \to \Big({4 G_5\over \pi}\Big)\, q_i \;,\quad\quad S_{\rm BH}\to
  \Big({4 G_5\over \pi}\Big)\, S_{\rm BH}\;.
  \label{restoreg5}
 \ee
Clearly the $q_i$'s defined in this way will not be integers.

\section{Black holes at $T=0$}
\label{scomparison}

In this Appendix we show that the $AdS_2\times S^{d-2}$ geometries derived in the text
agree with those coming by taking the zero temperature limit of the most general
non-extremal black hole solutions in $d=4,5$ dimensions.
For simplicity we focus on the static case.
 We refer the reader to \cite{Duff:1999rk} for details and references on
 the AdS black hole solutions quoted in this Appendix.

\subsubsection*{$d=4$ case}

The general non-extremal and static asymptotically AdS black hole solution of
$U(1)^4$ gauged supergravity in $d=4$ can be written as\footnote{The $X_I$'s here
are the inverse of the $X_i$'s used in \cite{Duff:1999rk}}:
\bea
ds_4^2 &=& H^{-2}\, f\, dt^2+H^{2}\,
(f^{-1}\, dr^2+r^2\, d\Omega_2)\;,\nn\\
X_I &=&{H_I\over H}\;, \quad\quad~~~~~ F^I~=~ d H_I^{-1}\,\coth \beta_i dt
\;,
\eea
with
\be
f=1-{m\over r}+4 \,g^2\, r^2\, H^4\;,\quad\quad H^4=H_1 H_2 H_3 H_4\;,
\quad\quad H_I=1+{m\sinh^2 \beta_i\over r}
\;.
 \ee
 The parameters $\beta_i$ and $m$ parametrize the electric charges and mass of the black
 hole. For a generic choice of $m$ the black hole has two horizons
 at $r_{\pm}$ given by the zeros of $f$. The two horizons
coincide when $r_0=r_+=r_-$, i.e.
 when both $f$ and its first derivative vanish at $r=r_0$:
\be
f(r_0)=f'(r_0)=0\;. \label{to4}
\ee
 Denoting
 \bea
 \ft12\mu_I &=& r_0\,H_I(r_0)\;,\\[1ex]
  \gamma_1 &=& \sum_I \mu_I\;, \quad \gamma_2=\sum_{I<J} \mu_I \mu_J\;,\quad
  \gamma_3=\sum_{I<J<K} \mu_I \mu_J\mu_K \;,\quad  \gamma_4=\mu_1 \mu_2\mu_3 \mu_4\nn
\;,
\eea
 equations (\ref{to4}) can be solved for $m$ and $r_0$ in terms of $\mu_I$:
 \bea
 m=g\, \sqrt{\gamma_4+\ft14\,g^2 \gamma_3^2}\;,\quad\quad
 r_0=\ft12\,m-\ft14\, g^2\,\gamma_3\;.
 \eea
The temperature of the black hole is zero for this choice and the
 horizon geometry takes the $AdS_2\times S^2$ form with
 \bea
 v_1 &=&\ft12\,H(r_0)^2\, f''(r_0)^{-1} ={\ft14 \sqrt{\gamma_4}\over
 1+g^2\, \gamma_2}\;,\qquad
 v_2 ~=~ r_0^2 \, H(r_0)^2=\ft14 \sqrt{\gamma_4}
\;,
 \eea
 in precise agreement with (\ref{genq}).

\subsubsection*{$d=5$ case}

The general non-extremal and static asymptotically AdS black hole solution of
$U(1)^3$ gauged supergravity in $d=5$ dimensions can be written
as\footnote{The $X_I$'s here are the inverse of the $X_i$'s used in \cite{Duff:1999rk}}:
\bea
ds_4^2 &=& H^{-2}\, f\, dt^2+H\,
(f^{-1}\, dr^2+r^2\, d\Omega_2)\;,\nn\\
X_I &=&{H_I\over H}\;, \quad\quad~~~~~ F^I~=~ d H_I^{-1}\,\coth \beta_i dt
\;,
\eea
with
\be
f=1-{m\over r^2}+g^2\, r^2\, H^3\;,\quad\quad H^3=H_1 H_2 H_3\;,
\quad\quad H_I=1+{m\sinh^2 \beta_i\over r^2}\;.
 \ee
 The parameters $\beta_i$ and $m$ parametrize the electric charges and mass of the black
 hole. For a generic choice of $m$ the black hole has two horizons
 at $r_{\pm}$ given by the two positive zeros of $f$. The two horizons
 coincide $r_0=r_{\pm}$ when parameters are chosen such that both $f$ and its
 first derivative vanish at the horizon:
\be
f(r_0)=f'(r_0)=0 \;.\label{to5}
\ee
 Denoting
 \bea
 \mu_I &=& r_0^2\,H_I(r_0)\;,\qquad
  \gamma_1 = \sum_I \mu_I\;,
  \quad \gamma_2=\sum_{I<J} \mu_I \mu_J\;, \quad
  \gamma_3=\mu_1 \mu_2\mu_3\;, \nn
\eea
 equations (\ref{to5}) can be solved for $m$ and $r_0$ in terms of $\mu_I$:
 \bea
 m=g\, \sqrt{4\gamma_3+g^2 \gamma_2^2}\;,\quad\quad
 r_0^2=\ft12 m-\ft12\, g^2\,\gamma_2\;.
 \eea
 The temperature of the black hole is zero for this choice and the
 horizon geometry takes the $AdS_2\times S^3$ form with:
 \bea
 v_1 &=&\ft12\,r_0^4\, H(r_0)\, f''(r_0)^{-1} ={\ft14\gamma_3^{1\over 3}\over
 1+g^2\, \gamma_1}\;,\qquad
 v_2 ~=~ r_0^2 \, H(r_0)=\gamma_3^{1\over 3}
 \;,
 \eea
 in agreement with (\ref{sol5genm}).

\end{appendix}


\providecommand{\href}[2]{#2}\begingroup\raggedright\endgroup

\end{document}